\documentclass{osa-article}

\journal{oe}
\articletype{Research Article}
\usepackage{bm}% bold math
\usepackage{siunitx}
\newcommand{\pvec}[1]{\vec{#1}\mkern2mu\vphantom{#1}}
\usepackage{hyperref}% add hypertext capabilities
\begin{document}

\title{Multimode Single-Pass Spatio-temporal Squeezing}

\author{Luca La Volpe,\authormark{1,*} Syamsundar De,\authormark{1,2,$\dagger$} Tiphaine Kouadou,\authormark{1}, Dmitri Horoshko,\authormark{3,4}, Mikhail I. Kolobov,\authormark{3}, Claude Fabre,\authormark{1} Valentina Parigi,\authormark{1} and Nicolas Treps,\authormark{1}}

\address{\authormark{1}Laboratoire Kastler Brossel, Sorbonne Universit\'{e}, CNRS,ENS-PSL Research University, Coll\`{e}ge de France; 4 place Jussieu, 75252 Paris, France\\
\authormark{2}Integrated Quantum Optics Group, Applied Physics, 100 Warburger Stra$\beta$se, Paderborn University, Paderborn 33098, Germany\\
\authormark{3}Univ. Lille, CNRS, UMR 8523, PhLAM-Physique des Lasers Atomes et Mol\'{e}cules, F-59000 Lille, France\\
\authormark{4}B. I. Stepanov Insitute of Physics, NASB, Nezavisimosti Ave. 68, Minsk 220072, Belarus\\
\authormark{5}Applied Physics Department, Université de Genève, 22 chemin de Pinchat, 1211 Genève 4, Switzerland
\email{\authormark{$\dagger$}syamsundar.de@upb.de}}

\email{\authormark{*}Luca.LaVolpe@unige.ch} %% email address is required



\begin{abstract}
We present a single-pass source of broadband multimode squeezed light with potential application in quantum information and quantum metrology. The source is based on a type I parametric down-conversion (PDC) process inside a bulk nonlinear crystal in a non-collinear configuration. The generated squeezed light exhibits a spatiotemporal multimode behavior that is probed using a homodyne measurement with a local oscillator shaped both spatially and temporally. Finally we follow a covariance matrix based approach to reveal the distribution of the squeezing among several independent temporal and spatial modes. This unambiguously validates the multimode feature of our source.
\end{abstract}

%%%%%%%%%%%%%%%%%%%%%%%%%%  body  %%%%%%%%%%%%%%%%%%%%%%%%%%

\section{\label{sec:level1}Introduction}

Squeezed states, in which squeezing is distributed to a set of independent optical modes, constitute an important quantum resource in the field of continuous-variable quantum information technology \cite{Braunstein2005b}, as for example, one-way quantum computation \cite{Menicucci2006}, and quantum communication \cite{Furusawa1998}. Besides, multimode squeezed light is a promising tool in  metrological applications, in particular for multi-parameter estimation with quantum enhanced sensitivity \cite{Pinel2012,Gessner2018}. Examples include quantum imaging via spatially multimode squeezing \cite{Embrey2015,Kolobov2007}, and quantum improved synchronization of distant clocks exploiting temporal/spectral multimode squeezed light \cite{Lamine2008}. The aforementioned wide range of potential applications are closely linked to the ever-increasing ability to generate, control, and detect multimode quantum light, thanks to the development of optical technologies such as spatial light modulators, optical frequency combs, multi-pixel detectors, to name a few.  

Squeezed light is commonly obtained from parametric down conversion (PDC) in a second-order nonlinear crystal placed inside an optical cavity, a so-called optical parametric oscillator (OPO). The optical cavity enhances the nonlinear interaction as well as it confines the squeezed light to a single spatial mode. Multimode squeezing has been produced by exploiting different degrees of freedom of light such as temporal/spectral \cite{Roslund2013}, spatial \cite{Treps2003}, and orbital angular momentum \cite{Lassen2009}. However, the OPO resonator limits the squeezing bandwidth to the resonator bandwidth. A promising alternative to produce broadband multimode squeezing is a single-pass PDC source, pumped with a pulsed laser which features an optical frequency comb in the frequency domain \cite{Wasilewski2006}. The single-pass design with a pulsed pump ensures that squeezing sustains at every pulse of the PDC output \cite{Wenger2004, Slusher1987}. Nonlinear waveguide based single-pass PDC sources are interesting to obtain high level of squeezing due to tight confinement of light \cite{Eto2011}. However, waveguide-based designs traditionally suffer from high loss that deteriorates the purity of the generated squeezed state \cite{Lenzini2018}. Furthermore, most of the existing studies on single-pass PDC sources, to the best of our knowledge, only followed a single-mode approach to characterize the generated squeezed light.

%Moreover, the absence of any optical cavity provides the possibility to include multimode feature in spatial degrees of freedom. 

In this study, we report multimode analysis of a single-pass source of broadband squeezed light based on a type I PDC process inside a bulk nonlinear crystal, in a non-collinear configuration. This particular design of our source leads to the fact that the generated squeezed light exhibits multimode features in both spatial and temporal/spectral degrees of freedom. To characterize the spatiotemporal multimode feature, we implement a homodyne detection in which the local oscillator (LO) is shaped both spatially and temporally. Finally, we construct covariance matrices from the results of the homodyne measurement and perform eigenvalue decomposition of the measured covariance matrices, providing a definite proof for multimode squeezing.

\section{Theoretical model}
\label{sec:TheoMod}

A theoretical analysis of a single-pass pulsed optical parametric amplifier in the temporal domain is reported in \cite{Wasilewski2006,Patera2009}, where the temporal modes in a waveguide are considered. In our case, we employ a bulk crystal where the spatial modes are not fixed by the waveguide geometry and, therefore, spatiotemporal coupling effects  become important \cite{Gatti2009}. It is convenient to consider the fields in the Heisenberg picture and in the slowly varying envelope approximation:
\begin{equation}
\hat{E}^{(+)}_{s,p}\left(\vec{x},z,t\right)=iE_0\int\frac{\mathrm{d}\vec{q}}{2\pi}\frac{\mathrm{d}\Omega}{2\pi}\hat{a}_{s,p}\left(\vec{q},z,\Omega\right)e^{i\left(\vec{q}\cdot\vec{x}-\left(\omega_{s,p}+\Omega\right)t\right)}
\end{equation}
where the indices $s$ and $p$ indicate, respectively, the down converted field (oscillating at a central angular frequency of $\omega_s$) and the pump of the process (with a central angular frequency of $\omega_p=2\omega_s$). The creation operator can be written in the slowly varying envelope approximation: $\hat{a}_{s,p}\left(\vec{q},z,\Omega\right)=\hat{A}_{s,p}\left(\vec{q},z,\Omega\right)e^{ik_z(\vec{q},\Omega)}$ with $\hat{A}$ being the slowly varying envelope operator and $k_z(\vec{q},\Omega)=\sqrt{k(\omega_s+\Omega)-\left|\pvec{q}\right|^2}$ being the projection of the wave vector along the propagation axis $z$.
The coordinates of interest are $\Omega$, the detuning from the central frequency $\omega_s=\omega_p/2$, and $\vec{q}$, the transverse wave-vector that corresponds to the position vector $\vec{x}$ in the Fourier plane of a lens. Usually being an intense coherent beam, and in the small pump depletion limit, the pump can be treated classically and the operator can be replaced by its mean value, $\hat{A}_p(\vec{q},z,\Omega)\rightarrow A_p(\vec{q},z,\Omega)$. The gain of the process is proportional to the non-linear susceptibility $\chi^{(2)}$, to the pump peak amplitude $\alpha_p$, and to the length of the crystal $\ell_c$, $g\propto \chi^{(2)}\alpha_p\ell_c$. As shown for example in \cite{Caspani2010}, in the low gain regime we can write the amplifier input-output relation:
\begin{equation}
\label{eq:Amplifier}
\begin{aligned}
\hat{A}_{s}^{\mathrm{out}}(\vec{q},\Omega)&=\hat{A}_{s}^{\mathrm{in}}(\vec{q},\Omega)+\\
&g\int\frac{\mathrm{d}\pvec{q}'}{2\pi}\frac{\mathrm{d}\Omega'}{\sqrt{2\pi}}\mathcal{K}(\vec{q},\Omega,\pvec{q}',\Omega')\hat{A}_{s}^{\mathrm{in}\dagger}(\pvec{q}',\Omega')
\end{aligned}
\end{equation}
where the kernel $\mathcal{K}$ is given by:
\begin{equation}
\label{eq:kernel}
\mathcal{K}(\vec{q},\Omega,\pvec{q}',\Omega')=A_p(\vec{q}+\pvec{q}',\Omega+\Omega')\mathrm{sinc}\left(\frac{\Delta(\vec{q},\pvec{q}',\Omega,\Omega')\ell_c}{2}\right).   
\end{equation}
Here, $A_p$ is the Fourier transform of the spatiotemporal profile of the pump and $\Delta$ is the wavevector mismatch along the pump propagation direction $z$: 
\begin{equation}
\Delta(\vec{q},\pvec{q}',\Omega,\Omega')=k_{sz}(\vec{q},\Omega)+k_{sz}(\pvec{q}',\Omega')-k_{pz}(\vec{q}+\pvec{q}',\Omega+\Omega')
\end{equation}
In our case of negative uniaxial crystal, the signal experiences the ordinary index $n_\mathrm{o}$ while the pump encounters the extraordinary index $n_\mathrm{ext}$, which depends on its orientation with respect to the optical axis of the nonlinear crystal:
\begin{equation}
\begin{aligned}
    k_s(\Omega)&=n_\mathrm{o}(\omega_0+\Omega)(\omega_0+\Omega)/c\\ k_p(q_y,\Omega)&=n_\mathrm{ext}(\theta_0+q_y/k_p(0,0),2\omega_0+\Omega)(2\omega_0+\Omega)/c
    \end{aligned}
\end{equation}
where $\theta_0$ is the angle between the optical axis and the pump propagation axis in the perfect phase matching condition for the desired non-collinear angle.

Applying an eigenvalue decomposition to the matrix obtained by discretizing the kernel in equation \eqref{eq:kernel}, the description of the problem simplifies \cite{Bennink2002}, and if the kernel is complex a Takagi decomposition can be used \cite{Arzani2018}. Considering the initial field $\hat{A}_{s}^{\mathrm{in}}$ as a linear combination of the eigenmodes, $S_k(\vec{q},\Omega)$, of the integral operator related to the kernel:
\begin{equation}
\hat{A}_{s}^{\mathrm{in}}(\vec{q},\Omega)=\sum_kS_k(\vec{q},\Omega)\hat{a}_k,
\end{equation} 
we simplify the equation of the amplifier, Eq. \eqref{eq:Amplifier}, as follows:
\begin{equation}
\label{eq:Aout}
\hat{A}_{s}^{\mathrm{out}}=\sum_kS_k(\vec{q},\Omega)\left(\hat{a}_k+g\Lambda_k\hat{a}_k^{\dagger}\right)
\end{equation} 
where $\Lambda_k$ is the corresponding eigenvalue. The quadratures of the eigenmodes are then independently squeezed with a squeezing parameter proportional to $g\Lambda_k$:
\begin{equation}
\label{eq:QuadSqz}
\hat{X}^{\mathrm{out}}_k=\left(1+g\Lambda_k\right)\hat{X}^{\mathrm{in}}_k,\quad\hat{P}^{\mathrm{out}}_k=\left(1-g\Lambda_k\right)\hat{P}^{\mathrm{in}}_k
\end{equation}

A detailed simulation of the spatio-temporal modes of a single-pass squeezer is shown in \cite{LaVolpe2019}. In our experiment as detailed in the following sections, we attempt to probe the spatiotemporal eigenmodes, $S_k(\vec{q},\Omega)$, as well as the corresponding squeezing parameter $2g\Lambda_k$ of our single-pass source by employing a homodyne detection with a spatial and temporal shaping of the local oscillator. 

%\textbf{comment: not really sure if the theory part is at all necessary or not}

\section{Experimental set-up}
\label{sec:Exset-up}
\begin{figure}[htbp]
\centering\includegraphics[width=.7\linewidth]{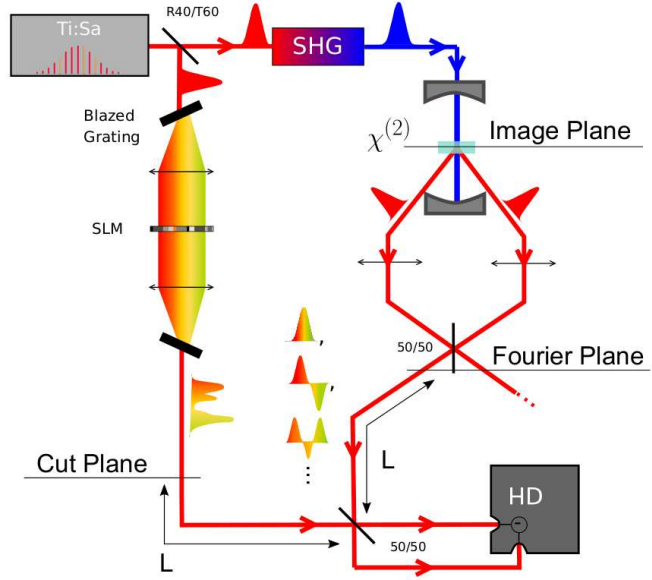}% Here is how to import EPS art
\caption{\label{fig:exset-up} Experimental set-up. In the figure we indicate the image plane and the Fourier plane for the signal and idler beams which are collimated by two identical lenses of \SI{200}{\milli\metre} focal length. The distance from the homodyne (HD) beam splitter and the Fourier plane corresponds to the distance of the cut plane from the HD beamsplitter in the LO path. }
\end{figure}

A scheme of the experimental set-up is shown in figure \ref{fig:exset-up}.
A titanium:sapphire (Ti:Sa) femtosecond oscillator, delivering \SI{22}{\femto\second} pulses at a repetition rate of \SI{156}{\mega\hertz}, constitutes the main frequency comb with a spectral FWHM of \SI{43}{\nano\metre} centred at \SI{795}{\nano\metre}. At the laser output, the pulses are separated by a beam splitter and 40$\%$ of the laser power is reserved for the LO while the rest is sent to the squeezer. A \SI{1}{\milli\metre} long bismuth-triborate (BiBO) crystal generates the pump for the PDC process, centered at a wavelength of \SI{397.5}{\nano\metre} with a spectral FWHM of \SI{1.82}{\nano\metre}, by frequency doubling the Ti:Sa pulses. The pump power is subsequently increased by a factor of seven using a synchronous cavity so that \SI{120}{\milli\watt} (0.77 nJ energy per pulse) are oscillating inside the linear cavity when it is locked with a Pound-Drever-Hall mechanism in transmission. The synchronous cavity also allows to clean the transverse profile of the pump beam.

The PDC crystal is placed at the waist (\SI{49}{\micro\metre}) of the cavity. We choose a \SI{2}{\milli\metre} long beta barium borate (BBO) as the nonlinear crystal which is set to optimize the type I PDC in the non-collinear configuration. The angle between the pump and the down converted beam is fixed at \ang{1.8} such that the down converted photons can be collected without passing through the cavity end-mirror to avoid losses. At the PDC output we get two beams which are then combined on a 50-50 beam splitter \cite{Wenger2005}. Finally, we implement a homodyne detection scheme in one of the beam splitter outputs to measure the quadratures. By shaping the temporal and spatial profiles of the LO we are able to analyse the amount of squeezing in each spatiotemporal mode and we can demonstrate the multimodal aspect of the source. 

%At this stage we have a multimode Einstein-Podolsky-Rosen (EPR) state which can be considered as a factorization of two-mode squeezed states. The shape of these modes consist in the sum and difference of the even and odd squeezed modes considered in equation \eqref{eq:Aout} \cite{Horoshko2019}.

%The two entangled pulses are then combined on a 50-50 beam splitter that leads to a squeezed state at each output of the beam splitter \cite{Wenger2005}. In order to perform a projective measurement of the squeezed light quadratures on the mode defined by the LO, one of the two outputs of the beam splitter is sent to a homodyne detection with a spectrally and spatially shaped LO. 

For the temporal shaping of the LO we use a pulse shaper based on a $792\times 600$ spatial light modulator (SLM) in a $4f$ line. A periodic saw-tooth grating is applied for each wavelength of the \SI{43}{\nano\metre} FWHM spectrum. The groove depth and position allow to control respectively the amplitude and phase for each spectral component \cite{Vaughan2005}. For the spatial shaping, we cut the beam at half power with a razor blade into left-cut (L) and right-cut modes (R), which are by construction orthogonal modes. As shown in figure \ref{fig:exset-up}, the distances from the homodyne beamspliter to the cut plane in the LO path and the Fourier plane of the squeezed beam are equal. 
In this way, the LO cut plane corresponds to the Fourier plane of the squeezed light beam.

%In this way, the LO-cut plane resembles the plane in the squeezed light beam where the transverse wavevectors are mapped into positions.

\section{Measurements}
Our first goal is to measure the amount of squeezing in different temporal/spectral as well as spatial modes. It is known that the temporal/spectral squeezed modes of the pulsed PDC process closely resemble Hermite-Gaussian (HG) modes if the spatiotemporal coupling is neglected  \cite{Wasilewski2006,Patera2009}. Therefore,  we implement different HG pulse shapes on the LO, thanks to the pulse-shaper in the LO path. We observe squeezing for the first four HG modes, where the first one, $\text{HG}_{0}$, has a spectral FWHM of \SI{15}{\nano\metre}. 
%As mentioned in the introduction, one of the most interesting features of a single-pass squeezing source is that squeezing can be measured on different spatial and temporal modes without acting on the source but only by shaping the LO. In our set-up this shaping is achieved with a pulse-shaper for the temporal domain and, for the spatial domain, by cutting the beam at the proper distance from the homodyne beam-splitter. We then chose two basis of orthogonal modes, one for the temporal domain and one for the spatial domain, and then observe the amount of noise reduction below the shot-noise level for each different mode.

\begin{figure}[htbp]
\centering\includegraphics[width=.7\linewidth]{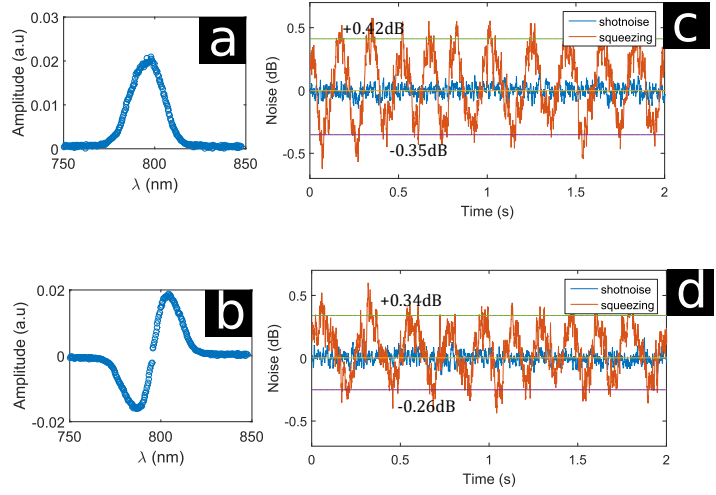}% Here is how to import EPS art
\caption{\label{fig:MMSqz} \textbf{a} and \textbf{b} Spectral amplitude of the first two Hermite-Gaussian (HG) modes as measured by a spectrometer. \textbf{c} and \textbf{d} variance of the homodyne signal as a function of time while sweeping the phase of the LO at \SI{300}{\milli\hertz} for, respectively, the 0th- and the 1st-order HG mode. The solid purple and the green line indicate the average value of squeezing and anti-squeezing, respectively.}
\end{figure}

%For the temporal domain we considered a four-mode Hermite-Gaussian (HG) basis, which is close to the eigenmodes of the amplifier if the spatio-temporal coupling is neglected \cite{Wasilewski2006}.
%By choosing the proper mask on the SLM we shaped the spectral phase and amplitude of the LO into each of the first four modes of the HG basis. This LO is then sent to the homodyne so that the quadrature noise relative to the chosen temporal/spectral mode can be measured. 
Figures \ref{fig:MMSqz}\textbf{a} and \ref{fig:MMSqz}\textbf{b} show the  spectral profile of the LO for the \text{0th}- ($\text{HG}_0$) and \text{1st}-order HG mode ($\text{HG}_1$), respectively. The traces in figures \ref{fig:MMSqz}\textbf{c} and \ref{fig:MMSqz}\textbf{d} report the noise variance of the homodyne signal, respectively, for $\text{HG}_0$ and $\text{HG}_1$ mode as a function of time while scanning the LO phase. The squeezing spectra are recorded with a spectrum analyser at a sideband frequency of \SI{10}{\mega\hertz} with a \SI{100}{\kilo\hertz} resolution bandwidth and \SI{30}{\hertz} video bandwidth, meaning that the measured squeezing is averaged over 15 pulses.
The amount of squeezing is \SI{-0.35\pm 0.03}{\decibel} in the $\text{HG}_0$ case, \SI{-0.25\pm 0.04}{\decibel} for $\text{HG}_1$, and it reduces to lower values for the higher order modes as expected from theory \cite{Horoshko2019}, so that we find \SI{-0.19\pm 0.05}{\decibel} for $\text{HG}_2$ and \SI{-0.19\pm 0.06}{\decibel} $\text{HG}_3$. The reason behind this low squeezing, compared to the existing studies of single-pass sources, is due to the limited pump pulse energy as well as not so tight confinement of the pump beam. 

%As can be seen, for the proper phase of the LO, the noise level goes below the shot-noise level. The latter is measured by blocking the light from the source and it is measured for each mode of the LO since their power is different. As from theory the first HG mode is more squeezed then the second. The amount of squeezing is low if compared with OPO based sources but, as mentioned before, in this case, because of the absence of a cavity, the same noise can be observed at any frequency up to the repetition rate of the laser, \SI{156}{\mega\hertz} in our case, the main technical limitation is the detector bandwidth. The higher number of squeezed modes in the single-pass configuration also reduces the squeezing per mode.  

\begin{figure}[htbp]
\centering\includegraphics[width=.7\linewidth]{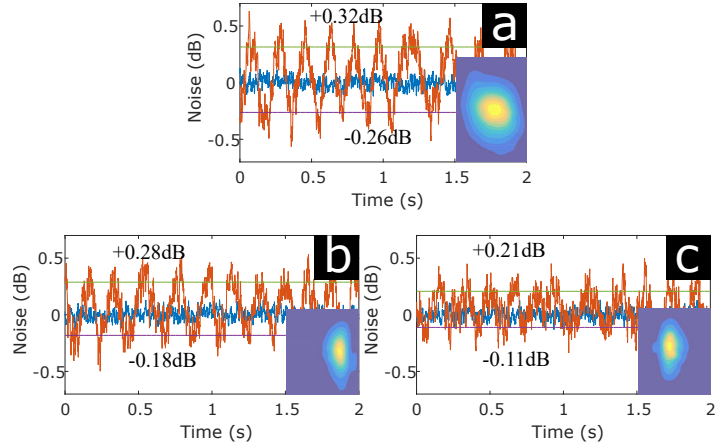}% Here is how to import EPS art
\caption{\label{fig:Razorcut} Variance of the homodyne signal versus time with a linear ramp of the LO phase for different spatial modes: \textbf{a} the whole beam, which can be seen as the sum of the other two modes, \textbf{b} the left-cut mode (L), and \textbf{c} the right-cut mode (R). The spatial shape of the mode, recorded with a CCD camera close to the homodyne beam splitter position, is shown in the inset of each sub-figure. The solid purple and the green line indicate the average value of squeezing and anti-squeezing,respectively.}
\end{figure}

In order to analyse multimode squeezing in the spatial domain, the LO beam is cut into two orthogonal modes, L and R. 
%we chose a simple basis of two orthogonal modes obtained by cutting the beam with a razor blade at exactly half power along the horizontal direction, parallel to the optical table (orthogonal to the pump polarization). As shown in figure \ref{fig:exset-up} the cutting plane is at a distance L from the homodyne beam splitter which matches the distance between the Fourier plane and the same homodyne beam splitter in the squeezed light path. In this way we know we are observing the plane were the transverse wave-vectors are mapped into positions. Since the beam is cut at half power, then the left-cut (L) and the right-cut (R) modes are orthogonal by construction and the whole beam is the sum of the two modes.
 The homodyne traces for the whole beam and the half-cut beams are shown in figure \ref{fig:Razorcut} with the corresponding spatial shape of the beam at the homodyne beam splitter in the insets. In these measurements, the $\text{HG}_0$ temporal mode is chosen for the LO. We find that both the half-cut modes are squeezed, \SI{-0.18\pm 0.04}{\decibel} for mode L and \SI{-0.11\pm 0.04}{\decibel} for mode R, though less compared to the whole beam, which is at \SI{-0.26\pm 0.04}{\decibel}.

%As can be seen, an amount of squeezing which is lower than that for the whole beam, is observed for both the orthogonal L and R spatial modes.
\subsection{Covariance matrix analysis}
Finally, we follow a covariance matrix approach to disclose the underlying independent squeezed modes, to unequivocally validate the multimode nature of the generated squeezed light \cite{Opatrany2002,Fabre2020}.  
%Even if we are able to measure squeezing in different modes, in general observing squeezing in different orthogonal spatial or temporal modes does not mean necessarily that the system under consideration is multimode. A clear and complete way to confirm this is a measurement of the covariance matrix describing the state \cite{Braunstein2005b}. 
The covariance matrix $\bm{V}$ is made of four blocks:
\begin{equation}
\bm{V}=\begin{bmatrix}
\bm{V^X}&\bm{V^{XP}}\\
\left(\bm{V^{XP}}\right)^T&\bm{V^P}
\end{bmatrix}
\end{equation}
where $V^X_{ij}=\langle\hat{X}_i\hat{X}_j\rangle$, $V^P_{ij}=\langle\hat{P}_i\hat{P}_j\rangle$ and $V^{XP}_{ij}=\frac{\langle\hat{X}_i\hat{P}_j\rangle+\langle\hat{P}_j\hat{X}_i\rangle}{2}$. Here $\hat{X}_i$ and $\hat{P}_i$ indicates the quadratures of the squeezed state related to the i-th mode of the LO basis. With this notation the diagonal elements of the covariance matrix report the variance of the X quadrature, $\langle\hat{X}_i^2\rangle$ or the P quadrature $\langle\hat{P}_i^2\rangle$ which are both equal to unity if the i-th mode is in a vacuum state. The off-diagonal elements represent the quantum correlations between the X or P quadratures related to different mode of the LO basis. A Bloch-Messiah decomposition of the covariance matrix gives a basis of eigenmodes whose quadratures are not correlated. If, in this basis, there exist more than one eigenvalue, which are different from 1 (the variance of the vacuum state), then the generated state of light can be genuinely labelled as a multimode \cite{Fabre2020}. 

Our measurement set up does not allow us to measure the correlations between the X and P quadratures of the modes i.e. the $\bm{V^{XP}}$ block. Correlations could be observed between the X and P quadratures if the pump was complex at the center of the crystal, however, in our case, the chirp acquired by the pump can be neglected and thus we can neglect this block of the covariance matrix. We therefore consider only the $\bm{V^X}$ (X block) and $\bm{V^P}$ (P block) and diagonalize them separately.

The quadrature variances of each squeezed mode are extracted from the corresponding homodyne trace. We relate the averaged maxima and minima of the homodyne trace to $\langle\hat{P}_i^2\rangle$ and  $\langle\hat{X}_i^2\rangle$, respectively. 
%Let us first consider the temporal domain. For each homodyne trace related to each HG mode we consider the average of the maxima to be related to the variance of the P quadrature, $\langle\hat{P}_i^2\rangle$ and the average of the minima to be related to the variance of the X quadrature $\langle\hat{X}_i^2\rangle$. 
To find the off diagonal elements of each block, we consider the following relation:
\begin{equation}
\langle\hat{X}_i\hat{X}_j\rangle=\frac{\langle\hat{X}_i+\hat{X}_j\rangle^2}{\sqrt{2}}-\frac{\langle\hat{X}_i\rangle^2}{2}-\frac{\langle\hat{X}_j\rangle^2}{2}
\end{equation}
where $\langle\hat{X}_i+\hat{X}_j\rangle^2$ is extracted from the homodyne measurement with a local oscillator in $(\text{HG}_i + \text{HG}_j)$ mode. 
In this way, all the elements of the X and the P block of the covariance matrix are measured.
%\textbf{Remark: Do we really need to put the full covariance matrix or simple $X_i$ block is enough?}

\subsubsection{Temporal domain}
\begin{figure}[htbp]
\centering\includegraphics[width=.7\linewidth]{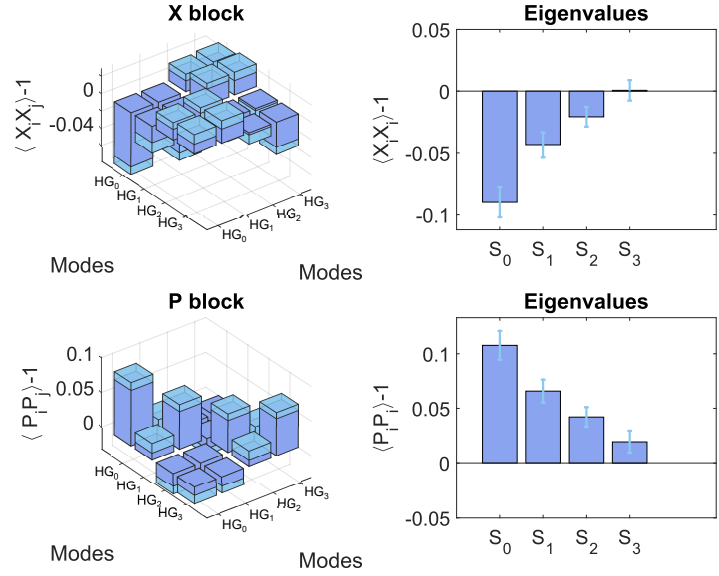}% Here is how to import EPS art
\caption{\label{fig:ExpCohTempo} Measured X and P blocks of the covariance matrix in the basis of the first four Hermite-Gaussian modes : $\lbrace\mathrm{HG}_0,\mathrm{HG}_1,\mathrm{HG}_2,\mathrm{HG}_3\rbrace$ in the temporal/spectral domain. On the right the eigenvalues for each block are reported, $S_0$, $S_1$, $S_2$ and $S_3$ are the relative eigenvectors.}
\end{figure}

We first report the results of the covariance matrix analysis in the temporal/spectral domain. In figure \ref{fig:ExpCohTempo} the X and the P block in the basis of the first four HG modes are shown on the left. For clarity, we remove the identity (the vacuum contribution) from the X and P blocks. These two matrices are independently diagonalized and their eigenvalues are shown on the right of figure \ref{fig:ExpCohTempo}. 

As shown in figure \ref{fig:ExpCohTempo}, there are more than one eigenvalues that have an absolute value higher than 1 for the P block and lower than 1 for the X block, validating the multimode feature of the generated squeezed light. As can be seen from the eigenvalues, their absolute value is higher for the P block eigenvalues (anti-squeezing) than for the X block eigenvalues (squeezing). This means that even in this diagonal basis the state is not pure. A reduced purity can be caused by losses in the experimental measurement or by an imperfect spatial or temporal overlap. Since in a single-pass squeezing source the losses can be neglected, then this reduced purity is probably due to a imperfect spatial overlap in this case. 

The large errors on the eigenvectors, unfortunately, do not allow us to infer the exact shape of the eigenmodes related to the eigenvalues. This is probably because the HG basis is close to the eigenmode basis, as can be seen from the X and P blocks in figure \ref{fig:ExpCohTempo}. As a result, the relative errors on the off-diagonal elements of the blocks become quite high, thus increasing the errors on the eigenvectors. Nevertheless, as can be seen from the right part of figure \ref{fig:ExpCohTempo}, the error bars on the eigenvalues are low enough to claim that the system is truly  multimode in temporal domain. 

\subsubsection{Spatial domain}

\begin{figure}[htbp]
\centering\includegraphics[width=1.0\linewidth]{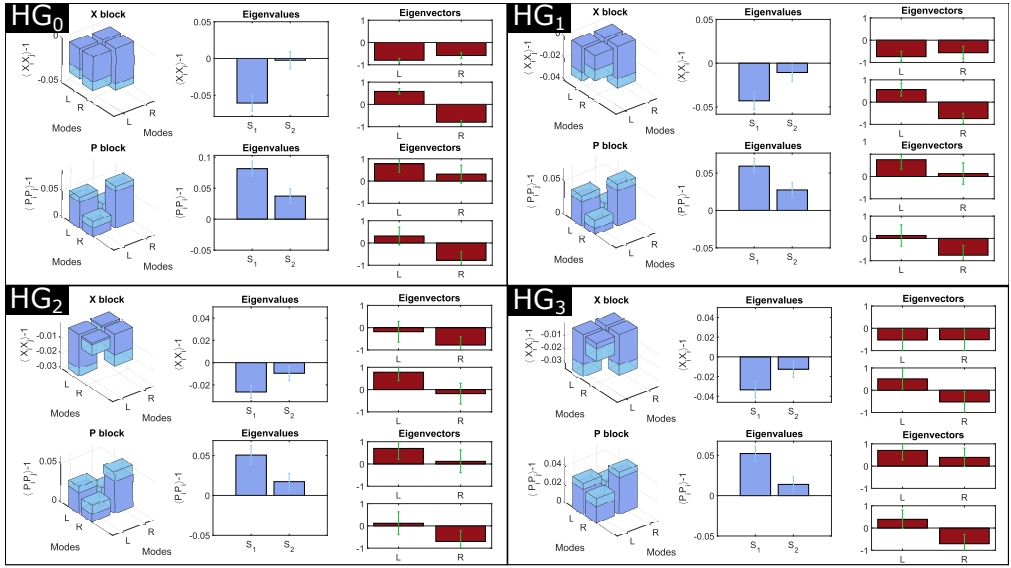}% Here is how to import EPS art
\caption{\label{fig:ExpCovSpace} Measurement of the X and P blocks of the covariance matrix in the basis of the left and right-cut modes: $\lbrace L,R\rbrace$ in the spatial domain. We report the X and P blocks together with the eigenvalues and eigenvectors in this spatial mode basis for each different Hermite-Gaussian temporal mode.}
\end{figure}

As in the temporal case, we analyse the spatial properties of our system through the covariance matrix with the basis of L and R modes. It is interesting to point out that we can record a covariance matrix in this spatial mode basis for each of the temporal mode that we used in the time domain analysis. We have then four different two-dimensional spatial covariance matrices in the L and R mode basis related to each temporal/spectral mode. In figure \ref{fig:ExpCovSpace}, we present the X and P blocks of the spatial covariance matrices related to each temporal mode $\text{HG}_i$. Right to the X and P blocks, the eigenvalues and eigenvectors for each matrix are reported.

For each of the temporal mode, we find that both the spatial modes are squeezed and, interestingly, the distribution of squeezing changes with the temporal modes. This is linked to the spatiotemporal coupling effect as explained in the theory section. To the best of our knowledge this is the first experimental measurement of the quadrature correlations in the spatial domain related to different time-frequency modes. The error bars on the eigenvectors are too big to infer the exact shape of the squeezed modes.  Nonetheless, for each temporal mode, we can see that the second eigenvector (the one related to the lowest eigenvalue) has always a $\pi$ phase shift between the left and the right part of the beam, while the first one does not. This indicates that the first squeezed mode in the spatial domain is a $TEM_{00}$ while the second is a flip-mode \cite{Treps2003}, and interestingly this feature is valid for every temporal/spectral mode. These measurements demonstrate that our source is multimode also in the spatial domain. However, it does not necessarily mean that the system has only two spatial squeezed modes. In fact, the number of detected eigenmodes is limited by the capabilities of our spatial mode shaping. In order to completely unveil the spatial distribution of squeezing, for example, a supplementary spatial-light modulator could be included in the LO path. 

%Furthermore, since for each temporal mode there is a two-mode squeezed state in the spatial domain, it should be possible to observe entanglement in the spatial domain for each temporal mode. Of course, in order to have a good signal-to-noise ratio in such a measurement, the amount of squeezing in the source should be increase in order to reduce the error bars on the squeezing measurements.

\section{Conclusion and perspective}
\label{sec:Perp}
In this study, we present a single-pass source of multimode squeezed light based on a non-collinear type I PDC process, pumped with an optical frequency comb. Though single-pass squeezing was already observed long ago, its multimodal nature was never experimentally demonstrated in the continuous variable domain. In this paper, we showed that we can measure the amount of squeezing in different temporal modes as well as in different spatial modes. Furthermore, we adopt a covariance matrix based approach to find the principal squeezed modes both in temporal and spatial domain, that serves up a clear signature of the multimodal behavior of our source. 
An interesting perspective for this source concerns the generation of a dual-rail cluster state, which has been shown to be a good candidate for quantum information\cite{Chen2014}. As clear from figure \ref{fig:exset-up}, the non-collinear configuration of our source allows, in principle, to put an inter-pulse delay between the signal and the idler pulses before recombining them on the beam splitter. The state generated in this way would be a large-scale dual-rail cluster state as reported in ref. \cite{Yokoyama2013}. Another exciting perspective of the present source with a seed beam is that it would be possible to generate bright beams, when seeding the PDC, with entangled temporal properties as demonstrated in the spatial domain \cite{Wagner2008}. 

\section*{Acknowledgments}
This  work  is  supported  by  the  French  National  Research  Agency  project SPOCQ and  the  European  Union  Grant  QCUMbER  (no.   665148). N.T. acknowledges financial support of the Institut Universitaire de France. V.P. acknowledges financial support from the European Research Council under the Consolidator Grant COQCOoN (Grant No. 820079).

Luca La Volpe and Syamsundar De contributed equally to this work.

%In this article we considered a promising configuration for the generation of multimode squeezed light which can be useful in metrology and quantum computing applications. Though single-pass squeezing was already observed long ago, its multimodal nature was never experimentally demonstrated. In the paper we showed we can measure the amount of squeezing related to different temporal modes, Hermite-Gaussian modes, and we measure the amount of squeezing in different spatial modes for four different HG modes in the temporal domain.

%Furthermore we properly demonstrated the multimode nature of the single-pass squeezing generation in the spatial and in the temporal domain by diagonalization of the covariance matrix.

%An interesting perspective for this source concerns the generation of  cluster states for quantum computation. As clear from figure \ref{fig:exset-up} the non-collinear configuration of the source allows in principle to put an inter-pulse delay between the signal and the idler pulses before recombining them on the beam splitter. The cluster state generated in this way is a dual rail cluster which has been shown to be a good candidate for quantum information\cite{Chen2014}. 

%%%%%%%%%% If using BibTeX:
\bibliography{sample}

%%%%%%%%%% If preparing manually:
% \newcommand{\enquote}[1]{``#1''}

% \bibitem{Zhang:14}
% Y.~Zhang, S.~Qiao, L.~Sun, Q.~W. Shi, W.~Huang, L.~Li, and Z.~Yang,
%   \enquote{Photoinduced active terahertz metamaterials with nanostructured
%   vanadium dioxide film deposited by sol-gel method,}
%   {\protect\JournalTitle{Optics Express}} \textbf{22}, 11070--11078 (2014).

% {Optical Society}, \enquote{{OSA Publishing},}
%   \url{http://www.osapublishing.org}.

% \bibitem{FORSTER2007}
% P.~Forster, V.~Ramaswamy, P.~Artaxo, T.~Bernsten, R.~Betts, D.~Fahey,
%   J.~Haywood, J.~Lean, D.~Lowe, G.~Myhre, J.~Nganga, R.~Prinn, G.~Raga,
%   M.~Schulz, and R.~V. Dorland, \enquote{Changes in atmospheric consituents and
%   in radiative forcing,} in \enquote{Climate Change 2007: The Physical Science
%   Basis. Contribution of Working Group 1 to the Fourth assesment report of
%   Intergovernmental Panel on Climate Change,}  S.~Solomon, D.~Qin, M.~Manning,
%   Z.~Chen, M.~Marquis, K.~B. Averyt, M.~Tignor, and H.~L. Miler, eds.
%   (Cambridge University Press, 2007).

% \end{thebibliography}

\end{document}